\documentclass[11pt]{article}
\usepackage{graphicx}
\usepackage{amssymb,amsthm,amsmath,float}
\usepackage{epstopdf}
\usepackage{hyperref}

\newcommand{\Eq}[1]{(\ref{#1})}
\newcommand{\Th}[1]{Th.~\ref{#1}}

\newcommand{\Sec}[1]{\S \ref{#1}}
\newcommand{\Fig}[1]{Fig.~\ref{#1}}

\newcommand{\InsertFig}[4]
{\begin{figure}[ht]
    \centerline{
     \includegraphics[width=#4]{./#1}
    }
    \caption{{\footnotesize #2}
    \label{#3}}
\end{figure}}

\newcommand{\R}{{\mathbb{ R}}}
\newcommand{\C}{{\mathbb{ C}}}

\newcommand{\cO}{{\cal O}}

\newcommand{\eps}{\varepsilon}
\newcommand{\vphi}{\varphi}

\newtheorem{tm}{Theorem}
\newtheorem{lm}[tm]{Lemma}

\newtheorem*{con}{Conjecture}

\title
{
  Chaotic dynamics of three-dimensional H\'enon maps that originate
  from a homoclinic bifurcation
}
\author
{
  S.V.Gonchenko\thanks
  {
    SVG and IIO were supported in part by grants of RFBR
    No.04-01-00487, No.04-01-00483 and No.05-01-00558; grant
    ''Russian Universities'' No. 03.01.180; grant NWO-RFBR No.
    047.017.018 and grant ''Leading scientific schools'' No.
    NSh-838.2003.2.
    JDM would like to acknowledge support of the NSF through grant
    DMS-0202032.
    This collaboration was supported by the CRDF under
    grant No.~RU-M1-2583-MO-04.
  }\\
    Inst. of Appl. Math. and Cybernetics\\ 10 Uljanova st.\\
    Nizhny Novgorod, 603005 Russia\\
    gosv100@uic.nnov.ru
  \and
  J.D. Meiss$^*$\\
    Applied Mathematics\\
    University of Colorado\\
    Boulder, CO 80309\\
    jdm@boulder.colorado.edu\\
  \and
  I.I.Ovsyannikov$^*$\\
    Nizhny Novgorod State Univ., Radio and Physical Dep.\\
    23 Gagarina av.\\
    Nizhny Novgorod, 603000, Russia\\
    ovii@mail.ru\\
}
\date{\today}

\begin{document}
\maketitle

\begin{abstract}
We study bifurcations of a three-dimensional diffeomorphism,
$g_0$, that has a quadratic homoclinic tangency to a saddle-focus
fixed point with multipliers $( \lambda e^{i\vphi}, \lambda
e^{-i\vphi}, \gamma)$, where $0<\lambda<1<|\gamma|$ and
$|\lambda^2\gamma|=1$. We show that in a three-parameter family,
$g_{\eps}$, of diffeomorphisms close to $g_0$, there exist
infinitely many open regions near $\eps =0$ where the
corresponding normal form of the first return map to a
neighborhood of a homoclinic point is a three-dimensional
H\'enon-like map. This map possesses, in some parameter regions, a
``wild-hyperbolic" Lorenz-type strange attractor. Thus, we show
that this homoclinic bifurcation leads to a strange attractor. We
also discuss the place that these three-dimensional H\'enon maps
occupy in the class of quadratic volume-preserving
diffeomorphisms.
\end{abstract}

\section{Introduction}\label{sec:intro}
In this paper we are concerned with the study of the
three-dimensional map $(\bar x, \bar y, \bar z) = f(x,y,z)\in
\R^3$ defined by
\begin{equation} \label{eq:3HM1}
  \begin{array}{l}
    \bar x=y,\;\; \bar y=z,\;\; \bar z=M_1+Bx+M_2y-z^2 \;,
  \end{array}
\end{equation}
with three parameters, $M_1$, $M_2$, and $B$. This map is a natural generalization
of the famous two-dimensional H\'enon map \cite{H76}; indeed, like the latter,
the map \Eq{eq:3HM1} is quadratic, has constant Jacobian $\det(Df) \equiv B$
and moreover, when $B=0$, reduces to the
two-dimensional H\'enon map. Therefore, it is natural to call \Eq{eq:3HM1}
the ``\textit{3D H\'enon map}." We will study a number of properties of
\Eq{eq:3HM1}, including:
 \begin{enumerate}
   \item its nontrivial dynamics and, most importantly, the existence of
      \textit{wild-hyperbolic} strange attractors;
  \item its origination in homoclinic bifurcations;
     \item its structure as a member of the family of
     three-dimensional quadratic maps with constant Jacobian.
\end{enumerate}

Three-dimensional maps have not been as widely studied as one- and
two-dimensional maps, though in recent years there has been a
considerable increase in their study. One of important reasons for
this fact is that multi-dimensional dynamical systems (the
dimension of the phase space is at least four for flows and three
for maps) can exhibit complicated dynamics that is decidedly
distinct from lower-dimensional cases. In particular they can
possess a new variety of strange attractor called a {\it
wild-hyperbolic attractor} \cite{ST98}.

Wild-hyperbolic attractors have no uniform hyperbolic structure;
moreover, they may contain homoclinic tangencies and coexisting
periodic orbits of different types. However, these attractors
contain no stable invariant subsets such as attracting periodic
orbits. Most crucially, the latter property holds for open regions
in the space of smooth dynamical systems. Thus, wild-hyperbolic
attractors are very interesting objects and their detection in
concrete models is a fundamental problem of nonlinear dynamics.

Most of the strange attractors occurring in two-dimensional, smooth maps (or
three-di\-men\-sional flows) are not structurally stable since they possess
homoclinic tangencies. For such systems, even though the orbits are observed
to be quite chaotic for some parameter values, very small changes in the parameters
may destroy the aperiodicity and give rise to periodic behavior. Such parameter
regions are called \textit{stability windows}. Even if these windows are not
``observable" in numerical computations, their appearance (under arbitrarily
small perturbations) is inevitable because of the prevalence of homoclinic tangencies.
 Such attractors were called \textit{quasi-attractors} in \cite{AS83}. One of the
most challenging problems in dynamical systems is to distinguish
aperiodic motion from extremely long periodic motion. Moreover,
since most of the chaotic attractors we see in applications are
definitely quasi-attractors, it is unclear in which sense they are
chaotic, or how one can define and measure the "probability" for a
quasi-attractor to be chaotic.

However, it is well-known that there are strange attractors that
are free of these problems; for example, hyperbolic
attractors and the Lorenz attractor. Though the latter is not
structurally stable \cite{GW79},\cite{ABS83} (in contrast to
hyperbolic attractors) it contains no stable periodic orbits,
every orbit in it has positive maximal Lyapunov exponent, etc.
Moreover, these properties are robust \cite{ABS77},\cite{GW79}.
The reason is that Lorenz attractor possesses a
pseudo-hy\-per\-bolic structure. In terms of Turaev and Shilnikov
\cite{ST98} this means that the following properties hold:
\begin{enumerate}
 \item there is a direction in which the flow is strongly contracting
	("strongly" means that any possible contraction in transverse
	directions is always strictly weaker); and
  \item transverse to this direction the flow expands areas.
\end{enumerate}
The robustness of pseudo-hyperbolicity ensures the robustness
of the chaotic behavior of orbits in the Lorenz attractor.
Moreover, pseudo-hyperbolicity is maintained even for (small, indeed)
time-periodic perturbations so that a periodically forced Lorenz
attractor is an example of a genuinely ``strange" attractor
since it contains no stable invariant subsets (for more
discussion see \cite{SST93},\cite{GOST05},\cite{TS05}). However it is
necessary to note that homoclinic tangencies arise inevitably
under such perturbations. Consequently systems with
periodically forced Lorenz attractors fall into Newhouse regions
and demonstrate an extremely rich dynamics \cite{GTS93},\cite{GTS99}.
Such attractors were called wild-hyperbolic
attractors in \cite{ST98};\footnote
{
	The term "wild" goes back to S.Newhouse
	\cite{N79} who introduced the term "wild hyperbolic set" for
	uniformly hyperbolic basic sets whose the stable and unstable
	manifolds (leaves) have a tangency. As Newhouse has proved, the
	latter property is persistent, consequently, systems with wild
	hyperbolic sets comprise open sets in the space of dynamical
	systems.
}
in this paper the theory of wild-pseudo-hyperbolic
attractor was discussed and an example of a wild-hyperbolic spiral
attractor was constructed.

Generalizing these ideas, we say that wild-hyperbolic
attractors possess the following two main properties that
distinguish them from other attractors: (1)
wild-hyperbolic attractors allow homoclinic tangencies (hence, 
they belong to Newhouse regions); (2) every such attractor and all nearby 
attractors (in the $C^r$-topology with $r\geq 2$) have
no stable periodic orbits.




One of the most felicitous properties of wild-hyperbolic
attractors is that they can be created in {\it local bifurcations}
of periodic orbits; in particular, when the orbit has three or
more Floquet multipliers on the unit circle \cite{SST93}.
Consequently, the construction of wild-hyperbolic attractors in
concrete models is possible when the model contains sufficiently
many parameters to permit such a degeneracy.  Indeed, the map
\Eq{eq:3HM1} is sufficiently complex and has been shown to possess
wild-hyperbolic attractors \cite{GOST05}.  Moreover, numerical
simulations have shown that these attractors persist in quite big
parameter domains and can be order-one in size \cite{GOST05}.

 In \Sec{sec:attractors} we will show that \Eq{eq:3HM1} has a small,
 wild-hyperbolic Lorenz attractor in a (small) parameter domain near the point
 $(M_1, M_2, B) = (-1/4, 1, 1)$ at which a fixed point with
 multipliers $(-1, -1,+1)$ exists.  The general case of fixed points
 with such a triplet of multipliers was considered in \cite{SST93},
 where it was shown that two distinct cases occur.  One of these cases
 has a local flow normal form that coincides with the Shimitsu-Marioka
 system (see \Eq{=sm}).  This system has a Lorenz attractor in some
 parameter domains \cite{ShA86},\cite{ShA93}.  As was noted in \cite{ST98} a
 periodically  forced Lorenz system (when the perturbation is rather small)
 can give  rise to a wild-hyperbolic attractor.  Moreover, this does not destroy the
 pseudo-hyperbolicity of the attractor \cite{ST98},\cite{GOST05}.  Note
 that the flow normal form only approximates the map and, since the
 residual terms are exponentially small in time, it is
 reasonable that the map will
 have a wild-hyperbolic attractor (which is small, in fact, because the
 approximation must be very good), see more discussions in
 \cite{GOST05}.

The second property of \Eq{eq:3HM1} that we mentioned above is that this map can arise as
a local normal form for multi-dimensional systems with a homoclinic tangency. This shows
that such maps are not exotic and gives a reason for an extensive study of their properties.

It is well-known that quadratic maps appear naturally when
studying bifurcations of quadratic homoclinic tangencies
\cite{GaS73I},\cite{GaS73II},\cite{TLY86},\cite{GTS93},\cite{GST03}.
They appear as normal forms of rescaled first return maps to the
neighborhood of a point on the homoclinic orbit.  There are four
standard forms for such one- and two-dimensional ``homoclinic
maps:"
\begin{enumerate}
  \item $\bar y = M_1 - y^2$ (the logistic map);
  \item $(\bar x, \bar y) = (y, M_1 - Bx - y^2)$ (the standard H\'enon map);
  \item $(\bar x, \bar y) = (y, M_1 + M_2 y - x^2)$ (the Mira map);
  \item $(\bar x,\bar y) = (y, M_1 - Bx - y^2 + \eps_k^1 xy + \eps_k^2 y^3)$
    (a generalized H\'enon map).
\end{enumerate}

The third case is a two-dimensional endomorphism and is one of the
well-studied polynomial maps (see, e.g. \cite{MGBC96}); we call it
the ``Mira map" following \cite{GST03}).  The fourth case can be
obtained as a small perturbation (due to $\eps_k^{1,2}$-terms) of
the standard H\'enon map.  This map was obtained in a study of
bifurcations of two-dimensional diffeomorphisms with a homoclinic
tangency to a saddle fixed point (with unit Jacobian)
\cite{GG00},\cite{GG04}.

The 3D H\'enon map \Eq{eq:3HM1} is absent from this list, and in
\Sec{sec:local}--\Sec{sec:rescaling} we will fill this gap.

In \Sec{sec:local} we study bifurcations of a three-dimensional
diffeomorphism with quadratic homoclinic tangency to a saddle-focus
fixed point.  For a three-dimensional map, such points can be of two
types: $(2,1)$ or $(1,2)$ (sometimes called \textit{type A} and
\textit{type B}, respectively).  Recall that $(k,l)$ saddle is a fixed
point with $\dim(W^s) = k$, $\dim(W^u) = l$, and $k+l = n$.  In our
case the two-dimensional manifold will be assumed to correspond to a
complex conjugate pair of multipliers.  We will consider the case
where the the product the absolute value of the multipliers, $P_0$ is
one.

The bifurcations in the general case, where $P_0 \neq 1$, were
studied in \cite{GST96},\cite{GST03}.  We will use their
``rescaling method" to study our case as well.  The essence of
this method is that a suitable change of coordinates and
parameters in a neighborhood of the homoclinic orbit is found to
bring the map into a standard form.  If the homoclinic tangency is
assumed to be quadratic the resulting map is also typically
quadratic.  For example, for a type $(2,1)$ saddle-focus with $P_0
< 1$, the rescaled first return maps coincide with the standard
H\'enon map up to asymptotically small terms (when the time of
return tend to infinity).  For a saddle-focus of type $(1,2)$, the
corresponding rescaled first return map is the Mira map. Note the
condition $P_0< 1$ leads to maps that are effectively
two-dimensional; consequently all three-dimensional volumes near
the fixed point are collapsed.  This is not the case when $P_0=1$
for the unperturbed map when three-dimensional rescaled maps can
appear a priori.  Indeed, we will show that when $P_0 = 1$ the map
\Eq{eq:3HM1} occurs as the normal form of the rescaled map for
type $(2,1)$ saddle-focus; and for the type $(1,2)$ case the
corresponding normal form is\footnote{
    Note that \Eq{eq:3HM2} is also the
    rescaled first return map for a four-dimensional diffeomorphism
    that has a quadratic homoclinic tangency to a saddle-focus fixed
    point with two pairs of complex conjugate leading multipliers
    \cite{GTS93},\cite{GST96},\cite{GST03}.
}

\begin{equation}\label{eq:3HM2}
    \begin{split}
        \bar x = y,\;\; \bar y = z, \;\;
        \bar z = M_1 + Bx + M_2 z - y^2,
    \end{split}
\end{equation}
Both \Eq{eq:3HM1} and \Eq{eq:3HM2} are quadratic
three-dimensional maps with constant Jacobian $J\equiv B$. It is
natural to call them the three-dimensional H\'enon maps of the \textit{first}
and \textit{second kind}, respectively.

The quadratic 3D H\'enon maps are elements of the group of polynomial
automorphisms.  The study of the dynamics of polynomial mappings has a
long history applied dynamics; for example, they often used in the
study of particle accelerators \cite{DA96}.  The study of
\textit{Cremona maps}, i.e., polynomial maps with constant Jacobian,
also has intrinsic mathematical interest \cite{En58}.  An interesting
mathematical problem is to obtain a normal form for an arbitrary
degree-$d$ Cremona map.  This problem is unsolved; the major
obstruction is that it is not known in general that such a map has a
polynomial inverse.  This is the content of the much studied,
``Jacobian conjecture":
\begin{con}[O.T. Keller (1939)]
    Let $f: \R^n \rightarrow \R^n$ be a
    Cremona map. Then $f$ is bijective and has a polynomial inverse.
\end{con}

\noindent
This conjecture is still open.

The set of polynomial maps with polynomial inverses is called the
\textit{affine Cremona group}.  For the case of the plane, the
structure of this group was obtained by Jung \cite{Ju42}.  Jung's
theorem states that planar polynomial diffeomorphisms are ``tame;''
that is, they can be written as a finite composition of ``elementary''
(or triangular) and affine maps.  More recently Friedland and Milnor
\cite{FM89} showed that any map in this group is either conjugate to a
composition of generalized H\'enon maps,
$$
            \bar x = y  \;, \;  \bar y = -B x + p(y) \;,
$$
where $p(y)$ is a polynomial and $B \ne 0$, or else the map is
dynamically trivial.

Not much is known about the structure of the affine Cremona group for
higher dimensions, though it was recently proved that the
three-dimensional Cremona map
\[
 \begin{split}
    \bar x &= x - 2y(xz + y^2 ) - z(xz + y^2 )^2 \;, \\
    \bar y &= y + z(xz + y^2 ) \;, \\
    \bar z &= z \;,
\end{split}
\]
is not tame \cite{SU04}.  Thus the normal form for the 3D affine
Cremona group must consist of more than tame maps.  However, it is
still possible that the nontame maps are dynamically trivial, like the
above example.

Another interesting problem concerns the approximation of smooth
diffeomorphisms by polynomial diffeomorphisms.  One result along this
direction is for the class of $C^\infty$ symplectic diffeomorphisms,
which, as Turaev has shown \cite{Tu03}, can be approximated by
compositions of the symplectic Cremona maps
$$
   \bar x = x_0 + y\;,\; \bar y =  -x + \nabla p(y) \;,
$$
where $x,y, x_0 \in \R^n$ and $p$ is a polynomial.  Related
approximation methods are used in the study of particle accelerators
\cite{DA96}, and results have also been obtained for analytic
symplectic maps \cite{Fo96}.

Some work has been done on finding normal forms for quadratic
symplectic and quadratic volume-preserving maps.  Moser showed
that every quadratic symplectic map in $\R^{2n}$ can be written as
the composition of an affine symplectic map and a symplectic shear
\cite{Mo94}.  For the four-dimensional case, he obtained a normal
form.  For the case of volume-preserving maps, it was shown in
\cite{LM98},\cite{LLM99} that any quadratic three-dimensional,
volume-preserving map whose inverse is also quadratic and whose
dynamics is non-trivial can be transformed (by affine
transformations) to the form
\begin{equation}\label{eq:3HM3}
  \begin{split}
    \bar x &= y,\;\; \bar y = z, \;\; \\
    \bar z &= M_1 +  Bx+ M_2^1 y + M_2^2 z + a y^2 + b yz + c z^2,
  \end{split}
\end{equation}
where $B = 1$, $|a + b + c|=1$ and one of the parameters $M_2^1$ or
$M_2^2$ can be eliminated (by a coordinate shift).  By analogy with
the planar results, there are two additional cases but they have
trivial dynamics (see \Sec{sec:quadratic}).  We also note that it is
not hard to construct three-dimensional quadratic diffeomorphisms that
do not fall into this classification because they have an inverse of
higher degree.  For example, the diffeomorphism
\[
   \bar x = x \;, \; \bar y = y + x^2 \;, \; \bar z = z + y^2 \;,
\]
has a quartic inverse.

Note that both of the maps \Eq{eq:3HM1} and \Eq{eq:3HM2} fall into the
form \Eq{eq:3HM3}.  Thus the results of \cite{LM98} show that when
$B=1$ both of the 3D H\'enon maps occur as normal forms for quadratic
elements of the affine Cremona group.  Moreover, even when $B \neq 1$
the map \Eq{eq:3HM2} is a normal form for a certain class of quadratic
maps \cite{Ta01}.  Tatjer showed that every quadratic map, $f$, for
which $f \circ f$ is also quadratic, but has no invariant linear
foliation is linearly conjugate to \Eq{eq:3HM2}.  In
\Sec{sec:statement}, we will generalize the results of \cite{LM98} to
show that \Eq{eq:3HM3} is the the normal form for quadratic
diffeomorphisms with quadratic inverses when $B \neq 1$ as well.

\section{Statement of Results}\label{sec:statement}

Consider a three-dimensional diffeomorphism $g_0 \in C^r(\R^3)$ for
some $r \geq 5$ that satisfies the following conditions (see
\Fig{fig:type21}):
\begin{enumerate}
\renewcommand{\labelenumi}{(\Alph{enumi})}
  \item $g_0$ has a type $(2,1)$ saddle-focus fixed point $O$ with multipliers
    $(\lambda_0 e^{ i\vphi_0}, \lambda_0 e^{- i\vphi_0}, \gamma_0)$ where
    $0<\lambda_0<1<|\gamma_0|$ (w.l.o.g.
    $0<\vphi_0<\pi$),;
  \item $P_0 \equiv |J_0| = |\lambda_0^2 \gamma_0|= 1$.
  \item The stable, $W^s$, and unstable, $W^u$,
    invariant manifolds of the fixed point $O$ have a quadratic homoclinic
    tangency at the points of a homoclinic orbit $\Gamma_0$.
\end{enumerate}

\InsertFig{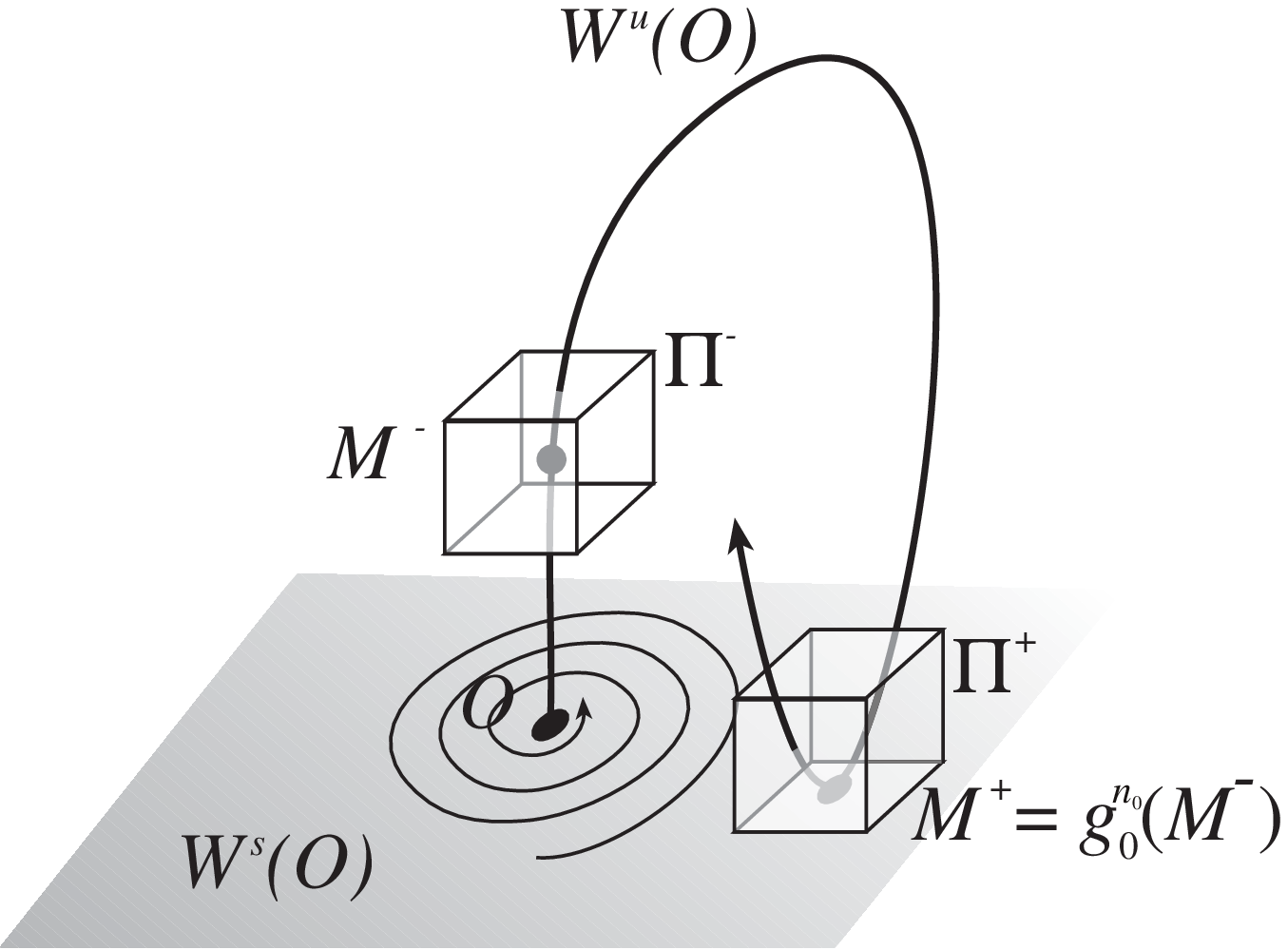}{Geometry of the homoclinic orbit to a type $(2,1)$
saddle-focus for the map $g_0$ under assumputions
(A)-(C).}{fig:type21}{3in}

Diffeomorphisms that are close to $g_0$ and satisfy conditions (B)
and (C) comprise, in the space of $C^r$
diffeomorphisms, a codimension-two, locally-connected surface
${\cal H}$---the bifurcation surface.

Our main problem is to study bifurcations of periodic orbits in
families of diffeomorphisms that transversely cross ${\cal H}$. We
denote such a family by $g_{\eps}$ so that $g_0$ satisfies
(A)-(C). The first question that arises is the optimal choice of
the governing parameters $\eps$. It is evident that the set of such
parameters must include both a parameter $\mu=\eps_1$ of the
splitting of the initial quadratic homoclinic tangency and a
parameter $\eps_2$, similar to $\eps_2= P_\eps-1$, that varies the
Jacobian of the saddle-focus fixed point. However, it was shown in
\cite{G00a} that diffeomorphisms on ${\cal H}$ possess continuous
(topological) conjugacy invariants on the set of non-wandering
orbits, so-called $\Omega$-moduli. The most important modulus is
the angular argument $\vphi$ of the complex multipliers of the
saddle-focus (see \Sec{sec:rescaling}). By definition,
$\Omega$-moduli are natural governing parameters. Therefore, we
include the $\Omega$-modulus $\vphi$ into the set of governing
parameters. Thus, we take as the third parameter $\eps_3
= \vphi - \vphi_0$.

More generally, we must consider a three-parameter family
$g_\eps$, $\eps=(\eps_1,\eps_2,\eps_3)$, that is transversal to
${\cal H}$ at $\eps =0$ and such that
\begin{equation*}
  \det \left(\frac{\partial(\eps_1,\eps_2,\eps_3)}
  {\partial(\mu,J,\vphi)}\right) \neq 0.
\end{equation*}

Consider a sufficiently small and fixed neighbourhood $U$ of the
orbit $\Gamma_0$. It can be represented as a union of a small
neighbourhood $U_0$ of $O$ with a number of small neighbourhoods
of those points of $\Gamma_0$ which do not belong to $U_0$. In the
family $g_\eps$ we study bifurcations of {\it single-round
periodic orbits} from $U$---that is orbits periodic that enter
$U_0$ exactly once. Every point of such an orbit of period $k$ can
be considered as a fixed point of the corresponding first return
map $g^k_\eps$ along orbits lying in $U$. In \Sec{sec:rescaling},
we will prove the following result about the form of this first
return map in some domains of the parameters.
\begin{tm}\label{tm:1}
  Let $g_\eps$ be the three parameter family defined above. Then,
  in any neighborhood of the origin, $\eps =0$, of the space of parameters, there
  exist infinitely many open domains, $\Delta_k$, such that the first return map
  $g^k_\eps$ takes the form
  \begin{equation}
    \label{eq:T07+1}
      \begin{array}{l}
        \bar X = Y + o(1)_{k\to +\infty}\;, \\ 
        \bar Y = Z + o(1)_{k\to +\infty}\;, \\ 
         \bar Z = M_1  + BX + M_2 Y - Z^2 + o(1)_{k\to +\infty} \;, \\ 
      \end{array}
  \end{equation}
  where $(X,Y,Z)$ are rescaled coordinates and the domains of definition of the new
  coordinates and the parameters $(M_1,M_2,B)$ are arbitrarily large and cover, as
  $k\to\infty$, all finite values. Furthermore, the Jacobian, $B$, is positive or negative
  depending on the orientability of the map $g^k_\eps$. The terms denoted $o(1)$
  are functions (of the coordinates and parameters) that tend to zero as $k\to +\infty$
  along with all derivatives up to order $r-2$.
\end{tm}

Theorem \ref{tm:1} shows that the 3D H\'enon map \Eq{eq:3HM1} is a normal form for this
homoclinic bifurcation, since it appears when we omit the $o(1)$-terms in \Eq{eq:T07+1}.
We will see that the new parameters $M_1,M_2$ and $B$ can be given as
functions of $\eps=(\mu, \varphi-\varphi_0,P_\eps-1)$ by
$$
    M_1\sim \gamma^{3k}(\mu + \cO(\lambda^k))\;,\;
    M_2\sim (\lambda\gamma)^k \cos (k\varphi + \beta)\;,\;
    B\sim (\lambda^2 \gamma)^k
$$
where $\beta$ is some coefficient (an invariant of $g_0$) and
$(\lambda(0),\gamma(0)) \equiv (\lambda_0,\gamma_0)$.  The exact
relations will be given in formulas \Eq{Mpar*}, \Eq{eq:T04} and
\Eq{eq:T06} below.  Thus, we can study the bifurcations of single
round periodic orbits in the family $g_\eps$, when $\eps \in
\Delta_k$, by studying the bifurcations of fixed points of the 3D
H\'enon map \Eq{eq:3HM1}.

Certainly, the problem of studying bifurcations in the 3D H\'enon
map is much simpler than the corresponding problem for the
original first return maps.  Moreover, the 3D H\'enon map provides
a simple, standard model for which various well-tried methods
(including numerical ones) can be applied.  However, the
bifurcation problem as whole for the 3D H\'enon map is very
complicated and here we focus only on two aspects. In
\Sec{sec:attractors} we first give the simple construction of
bifurcation surfaces, lines and points corresponding,
respectively, to codimension one, two and three bifurcations of
the fixed points.  Then we consider the problem of the existence
of wild-hyperbolic strange attractors. We will obtain the
following result.
\begin{tm}\label{tm:2}
  Let $g_\eps$ be the three parameter general family defined as before. Then,
  in any neighborhood of the origin $\eps =0$ of the space of parameters, there
  exist infinitely many open domains $\delta_k\subset\Delta_k$ such that the
  diffeomorphism $g_\eps$ of the family at $\eps\in\delta_k$ has a wild-hyperbolic strange
  attractor of Lorenz-type.
\end{tm}

The proof of this theorem is given in \Sec{sec:attractors}
following the analysis of \cite{GOST05}.  The main idea of the
proof is to apply the results of paper \cite{SST93} where it was
shown that maps having a fixed point with the triplet $(-1,-1,+1)$
of multipliers can, in some cases, be reduced to a flow normal
form that coincides with the Shimitsu-Marioka system (see
\Eq{fl1}) which possesses a Lorenz attractor for some parameter
domains \cite{ShA86},\cite{ShA93}.  Our verification of these
facts for the corresponding fixed point of the 3D H\'enon map
allows us to obtain \Th{tm:2}.  A justification of the fact that
the attractor in the 3D H\'enon map is actually the
wild-hyperbolic Lorenz-type attractor is given in \cite{GOST05}.
Consequently, we will omit these details.

Solving the bifurcation problem for the family $g_\eps$ allows us to
solve almost automatically an analogous problem for a diffeomorphism
$\tilde g_0 \in C^r$ ($r \ge 5)$ with a type $(1,2)$ saddle focus.
Specifically, assume that $\tilde g_0$ satisfies (see
\Fig{fig:type12}):

\begin{enumerate}
\renewcommand{\labelenumi}{(\Alph{enumi}$'$)}
   \item $\tilde g_0$ has a type $(1,2)$ fixed point $\tilde O$ with multipliers
    $(\lambda_0, \gamma_0 e^{i\psi_0}, \gamma_0 e^{-i\psi_0})$ where
      $0<|\lambda_0|<1<\gamma_0$ (w.l.o.g.~$0<\psi_0<\pi$);
   \item $P_0 \equiv |\lambda_0\gamma_0^2| = 1$; and
   \item the stable and unstable
    invariant manifolds of $\tilde O$ have a quadratic homoclinic tangency at
    the points of some homoclinic orbit $\tilde\Gamma_0$.
\end{enumerate}

\InsertFig{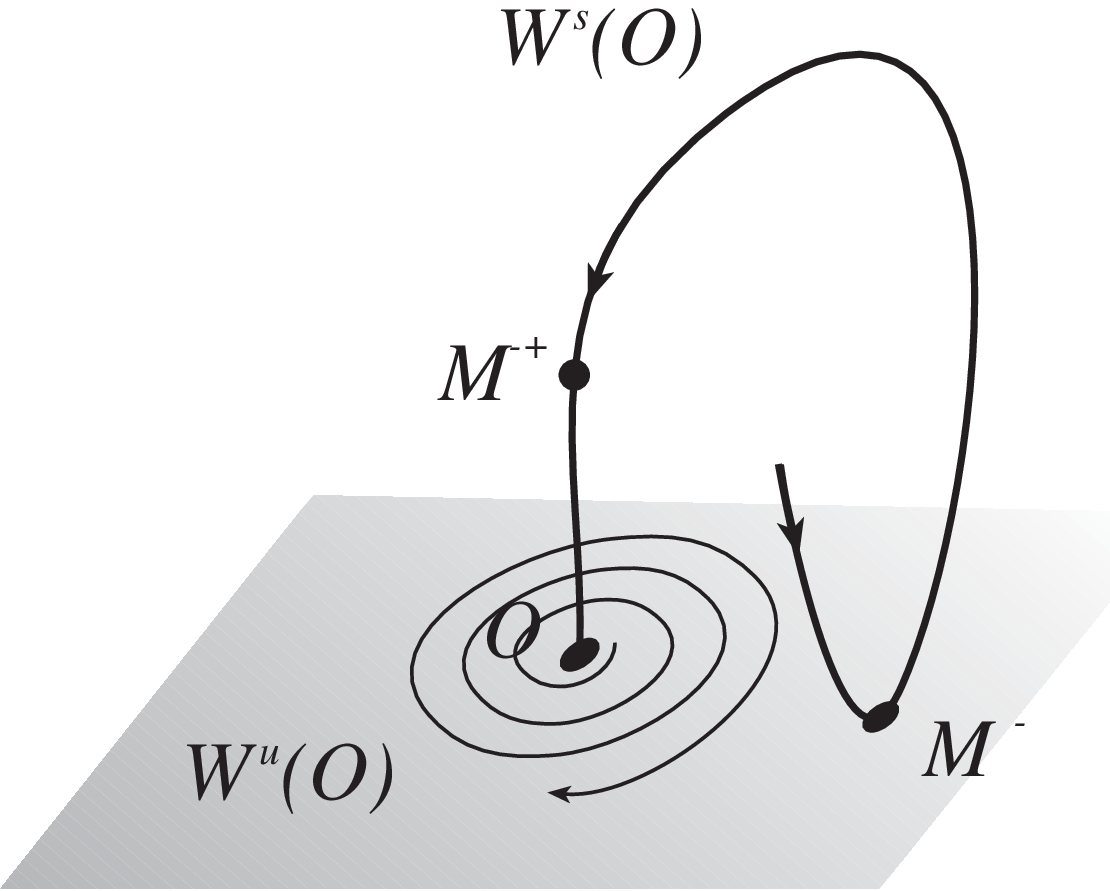}{Geometry of the homoclinic orbit to a type
$(1,2)$ saddle-focus for the map $\tilde g_0$ under assumptions
(A$'$)-(C$'$).}{fig:type12}{3in}
As in the case of the type $(2,1)$ saddle-focus, we consider a
three-parameter family $\tilde g_{\tilde\eps}$ where $\tilde\eps =
(\tilde\mu ,P_{\tilde\eps} -1, \psi - \psi_0)$.  We suppose here that
$\tilde\mu$ is a splitting parameter, $P_{\tilde\eps} =
|\lambda(\tilde\eps)\gamma(\tilde\eps)^2|$ and $\lambda(0)=
\lambda_0,\gamma(0)= \gamma_0$.

\begin{tm}\label{tm:3}
  Let $\tilde g_{\tilde\eps}$ be the three-parameter general family defined above. Then,
  in any neighborhood of $\tilde\eps =0$ there
  exist infinitely many open domains, $\tilde\Delta_k$, such that the first return map
  $\tilde g^k_{\tilde\eps}$ takes the form
  \begin{equation} \label{eq:T07+2}
      \begin{array}{l}
        \bar X = Y + o(1)_{k\to +\infty}, \\ 
        \bar Y = Z + o(1)_{k\to +\infty}, \\ 
        \bar Z = M_1 + BX + M_2 Z - Y^2 + o(1)_{k\to +\infty}, \\ 
      \end{array}
  \end{equation}
  where $(X,Y,Z)$ are rescaled coordinates and the domains of definition of the
  new coordinates and the parameters $(M_1,M_2,B)$ are arbitrarily large and cover,
  as $k\to\infty$, all finite values. Furthermore, the Jacobian, $B$, is positive
  or negative depending on the orientability of the map $\tilde g^k_\eps$.The terms
  denoted $o(1)$ are functions (of the coordinates and parameters) that tend to zero
  as $k\to +\infty$ along with all derivatives up to order $r-2$.
\end{tm}
\begin{proof} The proof immediate from \Th{tm:2}. Indeed, the
inverse $\tilde g_0^{-1}$ is, evidently, a diffeomorphism
satisfying the original conditions (A--(C). Thus, the
map \Eq{eq:T07+1} is, in this case, the rescaled map for the
inverse first return map $\tilde g_{\tilde\eps}^{-k}$. The
last can be be easily written in form (\ref{eq:T07+2}), as we show explicitly below.
\end{proof}

The inverse of \Eq{eq:3HM1} is of interest in any case. Since
\Eq{eq:3HM1} has the constant Jacobian $B$ it is invertible
whenever $B\neq 0$. This inverse is
\[ \begin{split}
    \bar y&= x , \;\; \bar z = y \;, \\
    \bar x &= \frac{1}{B}( -M_1 + z -M_2 y + y^2) \;.
  \end{split}
\]
This map can be rewritten in the ``standard'' form of \cite{LM98}, if
we define new variables $(x^{new}, y^{new},z^{new}) = B^{-1}(z,y,x)$ and new parameters
$$
  M_1^{new} = \frac{M_1}{B^2} \;,\;
  M_2^{new} = -\frac{M_2}{B}, \; B^{new} = \frac{1}{B} \;,
$$
then the resulting map has precisely the form \Eq{eq:3HM2}.  As we
stated above map is well-known in homoclinic dynamics: it was
found to be the rescaled first return map for four-dimensional
diffeomorphisms that have a quadratic homoclinic tangency to a
saddle-focus fixed point with two pairs of complex conjugate
leading multipliers \cite{GTS93},\cite{GST96},\cite{GST03}.

The dynamics of each of the maps \Eq{eq:3HM1} and \Eq{eq:3HM2} is
extremely rich.  However, for the volume-contracting case, we note
that these maps demonstrate distinct types of stable dynamics.
For example, \Eq{eq:3HM1} gives rise to attractors similar to
those of the Lorenz system, as we will see in
\Sec{sec:attractors}.  However, \Eq{eq:3HM2} need not have
attractors of this ``wild" type, although certainly, a ``Lorenz
repeller'' will be present.

Nevertheless, we can expect existence of ``extremely small"
Lorenz-type attractors in \Eq{eq:3HM2} by the following argument.
As we will see in \Sec{sec:attractors}, the map \Eq{eq:3HM2} can
have a fixed point with multipliers $(+1,e^{i\omega},
e^{-i\omega})$ at certain values of its parameters. The flow
normal form near this point can have a homoclinic loop to a
saddle-focus and, thus, \Eq{eq:3HM2} can have a fixed point $O$
with multipliers $(\gamma,\lambda e^{i\tilde\omega}, \lambda
e^{-i\tilde\omega})$, where $0<\lambda<1<\gamma$ and
$\lambda^2\gamma =1$, such that the manifolds $W^s(O)$ and
$W^u(O)$ will have a (quadratic) homoclinic tangency.
Consequently, we can apply \Th{tm:2} to this situation giving
immediately the promised ``extremely small" attractors.

Nevertheless, as it is shown in \cite{GOST05} the Lorenz-type
wild-hyperbolic attractor for \Eq{eq:3HM1} can be quite big and
were observed to have a large basin.  The proof of existence of big,
globally attracting wild-hyperbolic attractors in map \Eq{eq:3HM2}
is still open problem.

Finally, we consider the more general structure of polynomial
diffeomorphisms of $\R^3$.  As we remarked in the introduction, the 3D
H\'enon maps \Eq{eq:3HM1} and \Eq{eq:3HM2} have the form of
\Eq{eq:3HM3} when $B=1$.  As we will shown in \Sec{sec:quadratic},
this result has the following generalization;

\begin{tm}\label{tm:4}
    Let $f:\R^3\to\R^3$ be a quadratic diffeomorphism with a quadratic
    inverse and constant Jacobian $\det Df(x,y,z) = B$. Then, after an
    affine coordinate transformation $f$ can be written in the form \Eq{eq:3HM3}
    or else the dynamics of $f$ is trivial.
\end{tm}

Note that both of the maps \Eq{eq:3HM1} and \Eq{eq:3HM2} fall into
this classification. The classification of the dynamics of the
trivial cases can be done from the normal forms \Eq{eq:qforms}
that will be obtained in \Sec{sec:quadratic}.

\section{Local and global maps}\label{sec:local}

We begin the proof of \Th{tm:1} by constructing a local map $T_0$, valid in a
neighborhood $U_0$ of the fixed point $ O$, that simplifies the
description of the local unstable $W^u_{loc}(O) = W^u(O) \cap U_0$
and local stable $W^s_{loc}(O)= W^s(O) \cap U$ manifolds of $O$.
Then we obtain a global map $T_1$ that takes a neighborhood of a
point on $W^u_{loc}(O)$ to a neighborhood of a point of
$W^s_{loc}(O)$.

By \cite{GS93},\cite{SSTC98}, one can introduce $C^{r-1}$
coordinates $(x,y) = (x_1,x_2,y)$ in a neighborhood $U$ of the
fixed point $O$ such that the local map $T_0(\eps)\equiv
g_{\eps|U_0}$ can be written, for small enough $\eps$, in the
form\footnote {
    It follows from \cite{GS93},\cite{SSTC98} that \Eq{eq:T0} holds for
    general case, including $P_0 = 1$. For the
    volume-preserving case the analogous form was established in
    \cite{Baz93}}

\begin{equation} \label{eq:T0}
  T_0:
  \begin{array}{lll}
      \bar x & = &\lambda R_\vphi x + \xi(x,y)\\
      \bar y & =& \gamma y + \eta(x,y)
    \end{array}
\end{equation}
where
\begin{equation}\label{eq:rotation}
  R_\vphi = \begin{pmatrix} \cos\vphi & -\sin\vphi \\ \sin\vphi & \cos\vphi \end{pmatrix}
\end{equation}
is the rotation by angle $\vphi$, and
\begin{equation}\label{eq:xi_eta}
  \|\xi\| = \cO(\|x\|^2 y)\;, \; \mbox{and} \; \eta = \cO(\|x\| y^2) \;.
\end{equation}
Here the coefficients, $ \lambda,\vphi$, and $\gamma$ as well as
$\xi$ and $\eta$ depend (smoothly) on the parameters $\eps$. The
form \Eq{eq:T0} will be called the \textit{main normal form}; it
is very convenient for calculations. In particular, the local
stable and unstable manifolds of $O$ are $W^s_{loc} = \{ (x,0) \}$
and $W^u_{loc} = \{ (0,y) \}$ in the neighborhood $U$.

When $\eps= 0$, choose a homoclinic point
$M^- = (0,0,y^-) \in W^u_{loc}$ on the homoclinic
orbit $\Gamma_0$ and a corresponding point
$M^+ = (x_1^+,x_2^+,0)\in W^s_{loc}$ that is an
image of $M^-$: $M^+ = g_0^{n_0}(M^-)$ for some $n_0 > 0$, recall \Fig{fig:type21}.
Let $\Pi^+$ and $\Pi^-$ be some small neighborhoods of the points
$M^+$ and $M^-$. The global map
$T_1(\eps)\equiv {g_\eps^{n_0}}_{|\Pi^-}: \Pi^-\to\Pi^+$
is well defined for small enough $\eps$,
and can be written in the coordinates
of \Eq{eq:T0} as
\begin{equation}\label{eq:T1}
  T_1 :
  \begin{array}{ccl}
    \bar x-x^+ &=& A x + b (y-y^-) +
    \cO(\|x\|^2 + |y-y^-|^2), \\
    \bar y &=& y^+(\eps) + c^T x + d (y-y^-)^2 +
    \cO(\|x\|^2 + \|x\||y-y^-| + |y-y^-|^3),
  \end{array}
\end{equation}
where the $2 \times 2$ matrix $A$, the vectors $x^+$, $b$, and
$c$, the coefficients $y^+$ and $d$, and the higher order terms
all depend smoothly on $\eps$.  We noted especially the dependence
of the coefficient $y^+(\eps)$ that will become the splitting
parameter for the manifolds $W^s$ and $W^u$ near the homoclinic
point $M^+$.  As discussed in the previous section, we therefore
denote it by
\begin{equation} \label{eq:Van3}
  \mu = y^+(\eps)
\end{equation}
and consider as the first of the governing parameters.

The assumption (C) that the tangency between $W^u$ and $W^s$ is quadratic implies that
\begin{equation}\label{eq:nonzero_d}
  d \neq 0 \;.
\end{equation}
In addition, since $g_0$ is assumed to be a diffeomorphism, and
since the Jacobian
\begin{equation}\label{eq:J1}
  DT_1(M^-) = \begin{pmatrix} A & b \\ {c^T } & 0 \end{pmatrix} \;,
\end{equation}\
then we must have $\| b \| \neq 0$.

Indeed if $b_2(\eps)$ is not identically zero, then one can use the linear coordinate change,
\begin{equation}\label{eq:coordinate_change}
  (x_{new}, y_{new}) = (R_\vartheta x, y) \;,
\end{equation}
where $R_\vartheta$, \Eq{eq:rotation}, is a rotation by an angle
$\vartheta(\eps)$, to transform the map \Eq{eq:T1} to a new map of
the same form such that the vector $b$ becomes
\begin{equation}\label{eq:b1}
   b^T = (b_1, 0) \;.
\end{equation}
Here the new matrix $A$, point $x^+$, and vector $c$ are modified from
the corresponding coefficients in the original \Eq{eq:T1}.

Now the nonvanishing of the determinant of the Jacobian \Eq{eq:J1} implies that
\begin{equation}\label{Van7}
  J_1\equiv \det DT_1(M^-) = b_1(a_{21}c_2 - a_{22}c_1) \neq 0 \;.
\end{equation}

One of merits of the main normal form \Eq{eq:T0} is that the
iterations $T_0^k : U_0\to U_0$ for any $k$ can be calculated
quite easily, namely, in a form close to the form when $T_0$ is
linear.  In particular for the linear case we have
\[
    x_k = \lambda^k R_{k \vphi} x_0 \;, \quad y_0 = \gamma^{-k} y_k \;.
\]
This is the so-called {\it cross-form} of map $T_0^k$. We can
write $T_0^k$ in an analogous cross-form when $T_0$ is in the form
\Eq{eq:T0}

\begin{lm}\label{lem2} 
For any positive integer $k$ and for any sufficiently small $\eps$ the
map $T_0^k(\eps) :\; (x_0,y_0)\;\to\; (x_k,y_k)$  
can be written in the cross-form
\begin{equation}\label{eq:T0kk}
\begin{split}
  x_k  &= \lambda^k R_{k \vphi} x_0 + \lambda^{2k}\xi_{k}(x_0,y_k,\eps)\;,\\
  y_0  &= \gamma^{-k} y_k         +\lambda^{k}\gamma^{-k} \eta_k(x_0,y_k,\eps) \;,
\end{split}\end{equation}
where the vector $\xi_{k}$ 
and the function $\eta_{k}$ 
are uniformly bounded along with all derivatives up to order $(r-2)$. In addition,
all of the derivatives of order $(r-1)$ for the functions $x_k$ and $y_0$ tend to $0$ as
$k \rightarrow \infty $~.
\end{lm}

\begin{proof}
Let $(x_i,y_i) \in U \;,\;i=0,...,k,$ denote the orbit: $(x_i,y_i)=T_{0}(x_{i-1},y_{i-1})$.
Then it is easy to see that the $j^{\mbox{th}}$
iterate of \Eq{eq:T0} is 
\begin{equation}\label{eq:3*}
\begin{split}
  x_{j} & = \lambda^j R_{j\vphi} x_{0} + \sum\limits_{s=0}^{j-1}
    \lambda^{j-s-1}R_{(j-s-1)\vphi}\xi(x_s,y_s,\eps) \;,\\
  y_j & = \; \gamma^{j-k} y_k - \sum\limits_{s=j}^{k-1}
    \gamma^{j-s-1} \eta(x_s,y_s,\eps) \;.
\end{split}\end{equation}

The solution of \Eq{eq:3*} for any given sufficiently small $x_0
,y_k$ and large $k$ (the so-called boundary value problem) can be
found by successive approximations \cite{AS73} if the zeroth
approximation is taken as the corresponding solution of the linear
problem. Using the fact that $|\lambda\gamma| > 1$, we obtain the
following estimates
$$
  \begin{array}{l}
    \|x_k - \lambda^k R_{k\vphi} x_0\| \leq L \sum\limits_{s=0}^{k-1}
    \lambda^{k-s-1}\lambda^{2s}|\gamma|^{s-k} \leq
    \tilde L \lambda^{2k}, \\
    |y_0 - \gamma^{-k} y_k| \leq L \sum\limits_{s=0}^{k-1}
    |\gamma|^{-s-1}\lambda^{s}|\gamma|^{2s-2k} \leq
    \tilde L \lambda^{k}
    |\gamma|^{-k} ,
  \end{array}
$$
where $L$ and $\tilde L$ are some positive constants estimating
the norms for derivatives of the functions $\xi$ and $\eta$ in the
domain $U$, from \Eq{eq:T0}. These inequalities evidently imply
the desired result for the map. Estimates for the derivatives can
be found analogously. General results of this form for near saddle
fixed point were given in \cite{GS93},\cite{SSTC98}. Here we give
sharper estimates that apply in this particular case.
\end{proof}

\section{Rescaled first return map: Proof of \Th{tm:1}}\label{sec:rescaling}

The main task of this section is to obtain an analytical form for the
first return map for diffeomorphisms of the family $g_\eps$ to a
neighborhood of the homoclinic point $M^+$.  To construct this map we
first find an image of the local map $T_0$ for which $T_0^k(\Pi^+)
\cap \Pi^- \neq \emptyset$.  This will be true for each $k \ge \bar k$
where $\bar k$ is some sufficiently large integer.  For each such $k$
there are strips $\sigma_k^+ \in \Pi^+$ and $\sigma^-_k \subset \Pi^-$
such that $T_0(\sigma_k^+) = \sigma_k^-$.  A first return map is
obtained by composing $T_0^k$ with the global map $T_1: \Pi^-
\rightarrow \Pi^+$, to obtain a map on a neighborhood of $M^+$, recall
\Fig{fig:type21}.  Thus the first return map will be of the form
\[
  T_k\equiv T_1T_0^k \;,
\]
Under the global map $T_1$ the strip $\sigma_k^+$ is transformed
into three-dimensional horse-shaped region $T_1(\sigma_k^+)$ that
has non-empty intersection with $\sigma_k^+$ when $\eps$ is small
enough. Thus, the map $T_k$ is a generalized, three-dimensional,
horseshoe map.  We will show that these horseshoe maps can possess
rich dynamics; for example, they can Lorenz-like, wild-hyperbolic
strange attractors. The goal is the prove the following lemma:

\begin{lm} [Rescaling Lemma.] \label{lm:rescaling}
For any sufficiently large $k$ such that
\begin{equation}\label{eq:T02}
  \begin{array}{l}
    A_{21}(k\vphi) \equiv a_{21}\cos k\vphi + a_{22}\sin k\vphi \neq 0
  \end{array}
\end{equation}
 the first return map
$T_k$ can be brought, by means of an affine transformation of
coordinates, to the following form
\begin{equation}\label{eq:T07}
  \begin{array}{l}
    \bar Y = Z + \cO(\lambda^k)\;, \\
    \bar X = Y + \cO(\lambda^k + |\lambda\gamma|^{-k}) \;, \\
    \bar Z = M_1 - Z^2 + b_1 \lambda\gamma^{k} (c_1\cos k\varphi +
    c_2\sin k\varphi + \nu_k^1) Y \\
    + b_1A_{21}\lambda^{2k}\gamma^{k}(c_2\cos k\varphi -
    c_1\sin k\varphi + \nu_k^2) X + \cO(\lambda^k) \;,
  \end{array}
\end{equation}
where $\nu_k^{1,2}=\cO(\lambda^k)$ are some coefficients,
\begin{equation}\label{Mpar*}
  \begin{array}{l}
    M_1 = -d \gamma^{2k}[\mu
    - \gamma^{-k}y^- +
    \lambda^k\left(\cos k\vphi(c_{1}x_1^+ c_{2}x_2^+) +
    \sin k\vphi (c_{2}x_1^+ - c_{2}x_2^+)\right) +
    \cO(\lambda^{2k})].
  \end{array}
\end{equation}
Also, in \Eq{eq:T07} the $\cO$-terms have the indicated
asymptotic behavior for all derivatives up to order $(r-2)$.
\end{lm}

\begin{proof}
Using formulas \Eq{eq:T1}, \Eq{eq:T0kk} and conditions
\Eq{eq:Van3} and \Eq{eq:b1} we can write the first return map
$T_k\equiv T_1T_0^k$ in the form
\begin{equation}\label{eq:Van8}
  \begin{array}{rll}
    \bar x_1&=& x_1^+ + \lambda^k(e_{11} x_1 + e_{12} x_2) + b_1 (y-y^-)
    + \alpha_1, \\
    \bar x_2&=& x_2^+ + \lambda^k(e_{21} x_1 + e_{22} x_2) +\alpha_2, \\
    \gamma^{-k} \bar y + \lambda^k\gamma^{-k}
    \eta_k(\bar x,\bar y,\eps) &=&
    \mu +  \lambda^k (c_{1}\cos k\vphi + c_{2}\sin k\vphi) x_1 + \\
    & & \lambda^k (c_{2}\cos k\vphi - c_{1}\sin k\vphi ) x_2 +
    d (y-y^-)^2 + \alpha_3,
  \end{array}
\end{equation}
where the $2 \times 2$ matrix $E = A R_{k \vphi}$ and
$$
  \begin{array}{l}
    \alpha_{1,2} = \cO(\lambda^{2k}\|x\| + \lambda^{k}\|x\||y-y^-| +
    (y-y^-)^2), \\
    \alpha_{3} = \cO(\lambda^{2k}\|x\| + \lambda^k\|x\||y-y^-|+
    |y-y^-|^3).
  \end{array}
$$
Introduce new coordinates
$$
  x_{new} = x-x^+ + \beta_k^1, \; y_{new} = y - y^+ \beta_k^2,
$$
where $\beta_k = \cO(\lambda^k)$, in order to eliminate the affine terms in
the first two equations and the term linear in $y$ in the third equation of \Eq{eq:Van8}.
Then, the map \Eq{eq:Van8} is rewritten as
\begin{equation}\label{Van9}
  \begin{array}{l}
    \bar x_1 = e_{11}\lambda^k x_1 + e_{12}\lambda^k x_2 + b y
    +
    \cO(\lambda^{2k}\|x\| + \lambda^k\|x\|\|y\|+ \|y\|^2), \\
    \bar x_2 = e_{21}\lambda^k x_1 + e_{22}\lambda^k x_2 +
    \cO(\lambda^{2k}\|x\| + \lambda^k\|x\|\|y\|+ \|y\|^2), \\
    \bar y + \lambda^k \cO(\|\bar x\| + \|\bar y\|) = M +
    d \gamma^k y^2 + \\
    + \lambda^k\gamma^k (c_{1}\cos k\vphi + c_{2}\sin k\vphi
    + \cO(\lambda^k))x_1 +
     \lambda^k\gamma^k(c_{2}\cos k\vphi - c_{1}\sin k\vphi
    + \cO(\lambda^k) )x_2 + \\
    + \gamma^k \cO(\lambda^{2k}\|x\|^2 + \lambda^k\|x\|\|y\|+ \|y\|^3),
  \end{array}
\end{equation}
where $d_{new}= d + \cO(\lambda^k)$ and
\begin{equation}\label{eq:T00}
  \begin{array}{rlr}
    M &=& \gamma^k [\mu
    - \gamma^{-k}y^- +
    \lambda^k\left(\cos k\vphi(c_{1}x_1^+ c_{2}x_2^+) +
    \sin k\vphi (c_{2}x_1^+ - c_{2}x_2^+)\right) +
    \cO(\lambda^{2k})].
   \end{array}
\end{equation}

 Next, we rescale the map \Eq{Van9} introducing new
coordinates (X,Y,Z) defined through
\begin{equation}\label{eq:T01}
    x_1 = - \frac{b_1}{d}\gamma^{-k} Y, \;\;
    x_2 = - \frac{b_1A_{21}}{d}\lambda^k\gamma^{-k} X, \;\;
    y = - \frac{1}{d}\gamma^{-k} Z,
\end{equation}
where we suppose that values of $\vphi$ are such that \Eq{eq:T02}
holds. Then, map \Eq{Van9} is rewritten in the new coordinates in
form \Eq{eq:T07}.
\end{proof}

Now theorem~\ref{tm:1} easily follows from this lemma. Indeed, we
have only to indicate those domains $\Delta_k$ of the initial
parameters where the coefficients in the third equation of
\Eq{eq:T07} are finite. Denote the coefficient of the third
equation of \Eq{eq:T07} in front of $Y$ as $M_2$, i.e.
\begin{equation}\label{eq:T04}
  \begin{array}{l}
    M_2 \equiv b_1\lambda^k\gamma^k(c_{1}\cos k\vphi + c_{2}\sin k\vphi +
    \nu_k^1).
  \end{array}
\end{equation}
Since $|\gamma|>1$ and $|\lambda\gamma|>1$, the coefficients $M_1$
and $M_2$ can take arbitrary finite values when varying parameters
$\mu$ (for $M_1$) and $\vphi$ (for $M_2$), see formulas \Eq{Mpar*}
and \Eq{eq:T04}. Moreover, the finiteness of $M_2$ means that the
corresponding values of the trigonometrical term $c_{1}\cos k\vphi
+ c_{2}\sin k\vphi$ are asymptotically small and, in any case, do
not exceed $|\lambda\gamma|^{-k}$ in order. For such values of
$\vphi$, we obtain
\begin{equation}
\label{eq:T05}
  \begin{array}{l}
    c_{2}\cos k\vphi - c_{1}\sin k\vphi = \pm \sqrt{c_1^2 + c_2^2}(1+...), \\
    \displaystyle A_{21}\equiv a_{21}\cos k\vphi + a_{22}\sin k\vphi =
    \pm \frac{a_{21}c_2 - a_{22}c_1}{\sqrt{c_1^2 + c_2^2}}(1+...)
  \end{array}
\end{equation}
Thus, by \Eq{Van7}, $A_{21}\neq 0$ in this case. Denote as $B$ the
coefficient in front of $X_2$ from the third equation of
\Eq{eq:T07}. By virtue of \Eq{eq:T05} and \Eq{Van7} we have
\begin{equation}\label{eq:T06}
  \begin{array}{l}
    B = b_1(a_{21}c_2 - a_{22}c_1) \lambda^{2k}\gamma^{k}(1+...) =
    J_1 \lambda^{2k}\gamma^{k}(1+...).
  \end{array}
\end{equation}
Since $|\lambda^{2}\gamma|=1$ at $\eps=0$, it follows from
\Eq{eq:T06} that $B$ can take  (when $k$ is large) any arbitrary finite
value (positive or negative depending on the sign of
$J_1\lambda^{2k}\gamma^{k}$)
as the value of the Jacobian $\lambda^{2}\gamma$ of the local map $T_0$ varies.
This completes the proof of theorem~\ref{tm:1}.

\subsection*{Remarks}

\begin{enumerate}
    \item We noted that at least three parameters must be used to
    study bifurcations of a homoclinic orbit for a
    three-dimensional saddle-focus. The governing parameters that we use
    are $\eps_1 = \mu$ corresponding to the splitting distance
    between the stable and unstable manifolds, the
    phase $\eps_2 = \vphi-\vphi_0$ of the focus multiplier,
    and the Jacobian $\eps_3 = |\lambda^2\gamma|-1$
    of the saddle-focus.

    \item If the map $g_0$ is exactly volume preserving,
    i.e., $\det Dg_0 \equiv 1$, then the resulting
    map \Eq{eq:T07} is as well, i.e. $B\equiv 1$.
    Such maps were studied in \cite{LM98}.
    Moreover, \Th{tm:2} implies that there
    are nearby maps (by unfolding the parameter $B$ to make the map volume contracting)
    that have strange attractors.
\end{enumerate}


\section{Local Bifurcations of 3D H\'enon maps}\label{sec:bifurcations}

In this section we will study the dynamics of the 3D H\'enon map
\Eq{eq:3HM1} and prove \Th{tm:2}.  The dynamics can be complex,
especially when $|B|$ is close to one, essentially because when
$|B|=1$ the map has, for certain values of the parameters, fixed
points with three multipliers are on the unit circle.  It is well
known that bifurcations of such points can lead to appearance of
homoclinic tangencies, invariant circles, and even strange
attractors (as $|B|$ decreases below one) or repellers (as $|B|$
increases above one) \cite{ShL81},\cite{SST93}.

We are interested primarily in the stable dynamics of the map
\Eq{eq:3HM1}, i.e., in characterizing its attractors.  One way to find
attractors with complex dynamics is to study the local normal forms
that arise near fixed points that have three multipliers that are
$\pm1$.  We follow here the analysis of \cite{SST93} where local
normal forms of the equilibrium of a flow with three zero eigenvalues
and with additional symmetries were analyzed and conditions for the
existence of strange attractors were given.\footnote
{   These results
    can be applied to the analysis of bifurcations of codimension-three fixed points of
    three-dimensional maps. In this case we can often construct some approximated
    flow normal form whose time-one shift map coincides (up to terms of a certain
    order) with some power of the original map restricted on a neighborhood of the
    fixed point. Usually, the presence of multipliers $-1$ or $e^{\pm i\psi}$
    implies that the corresponding approximated flow possesses some symmetry,
    such as a reflection $x\to -x$, or a rotation, $O_2$.
}
One of the most interesting of their results is that Lorenz
attractors can appear by local bifurcations of an equilibrium with
three zero eigenvalues when there is a reflection symmetry through
the origin \cite{ShL81},\cite{SST93},\cite{PST98}.  These results
can be applied to the local bifurcation analysis of \Eq{eq:3HM1}
when the fixed points has multipliers $(+1,-1,-1)$.

As is true for the general volume-preserving quadratic diffeomorphism
of Lomel\'i and Meiss \cite{LM98}, the 3D H\'enon map \Eq{eq:3HM1} has
at most two fixed points on the diagonal:
\begin{equation}\label{eq:fp}
\begin{split}
    P_\pm   &= (z_\pm^*,z_\pm^*,z_\pm^*) \;,\\
    z_{\pm}^* &=\frac12 \left({-1+B+M_2} \pm \sqrt{\Delta}\right) \;,
\end{split}\end{equation}
providing that the discriminant, $\Delta$, is nonnegative,
\begin{equation}\label{P1+}
     \Delta \equiv (M_2+B-1)^2+4M_1 \ge 0 \;.
\end{equation}

The primary bifurcations of these fixed points occur when they have
multipliers on the unit circle.  These bifurcations are easily
determined using the characteristic polynomial of the Jacobian of
\Eq{eq:3HM1},
\[
    p(\lambda) = \lambda^3 +2z\lambda^2 -M_2 \lambda - B \;.
\]
The first codimension-one case corresponds to a single multiplier one,
where $p(1) = 2z^*+1-B-M_2=0$, or equivalently to the surface where
$\Delta$ vanishes:
\begin{equation}\label{eq:sn}
        L^{sn}\;:\;\; M_1 = - \frac14 (M_2+B-1)^2 \;. \\
\end{equation}
This is simply a saddle-node bifurcation that creates the pair of
fixed points \Eq{eq:fp} in the domain $\Delta > 0$.  Similarly, a
period-doubling bifurcation occurs when there is a multiplier $-1$,
i.e., where $p(-1) = 2z^*-1-B+ M_2 = 0$, or equivalently for $1-M_2 =
\pm \frac12 \sqrt\Delta$, which gives
\begin{equation}\label{eq:pd}
        L^{pd}\;:\;\; M_1 = \frac14 (M_2-1-B)(3M_2-3+B) \;. \\
\end{equation}
Note that this bifurcation occurs for the point $P_+$ when $M_2 <1$,
and for the point $P_-$ when $M_2 > 1$.

The final codimension-one bifurcation corresponds to the Hopf bifurcation when
pair of multipliers is on the unit circle, $\lambda = e^{\pm i\vphi}$, so that
\[
    p(\lambda) = \lambda^3 -(B+2\cos\vphi)\lambda^2 +(1+2B\cos\vphi)\lambda -B \;,
\]
or equivalently on the surface
\begin{equation}\label{eq:hopf}
        L^{\varphi}\;:\;\;
    \left\{ \begin{array}{l}
      M_1 = (\cos\vphi + \frac12 B)\left((1-2B)\cos\vphi + \frac32 B -2\right),\\
            M_2 = -1 - 2B\cos\vphi \;. \\
    \end{array}\right.
\end{equation}
Here we have written the equation for the surface $L^{\varphi}$ in
parametric form using the phase $0 < \vphi < \pi$.  When $B<1$, this
bifurcation only occurs for the fixed point $P_+$.

There are three codimension-two curves in the $(M_1,M_2,B)$ parameter
space.  The Hopf surface, $L^h$ terminates at $L^{sn}$ when the
multipliers are $(B,1,1)$ along the curve $(M_1, M_2) = (\frac14
(B+2)^2 , -2B-1)$.  It also terminates on $L^{pd}$ where the
multipliers are $(B,-1,-1)$ along the curve $(M_1, M_2) = (\frac14
(B-2)(7B-6), 2B-1)$.  Finally, the period-doubling and saddle-node
curves are tangent when the multipliers are $(-B,-1,1)$ on the curve
$(M_1,M_2) = (-\frac14 B^2, 1)$.

The bifurcation surfaces are most easily viewed in two-dimensional
slices through constant $B$ planes.  An example of the resulting
bifurcation curves are shown when $B = 0.6$ in
\Fig{fig:type21bifB06fp1} and \Fig{fig:type21bifB06fp2}.  Note that
$P_+$ is stable inside the curvilinear triangle bounded by $L^{sn}$,
$L^{pd}$ and $L^{\varphi}$, when $B <1$.  The second fixed point,
$P_-$, is never stable when $B<1$.

\InsertFig{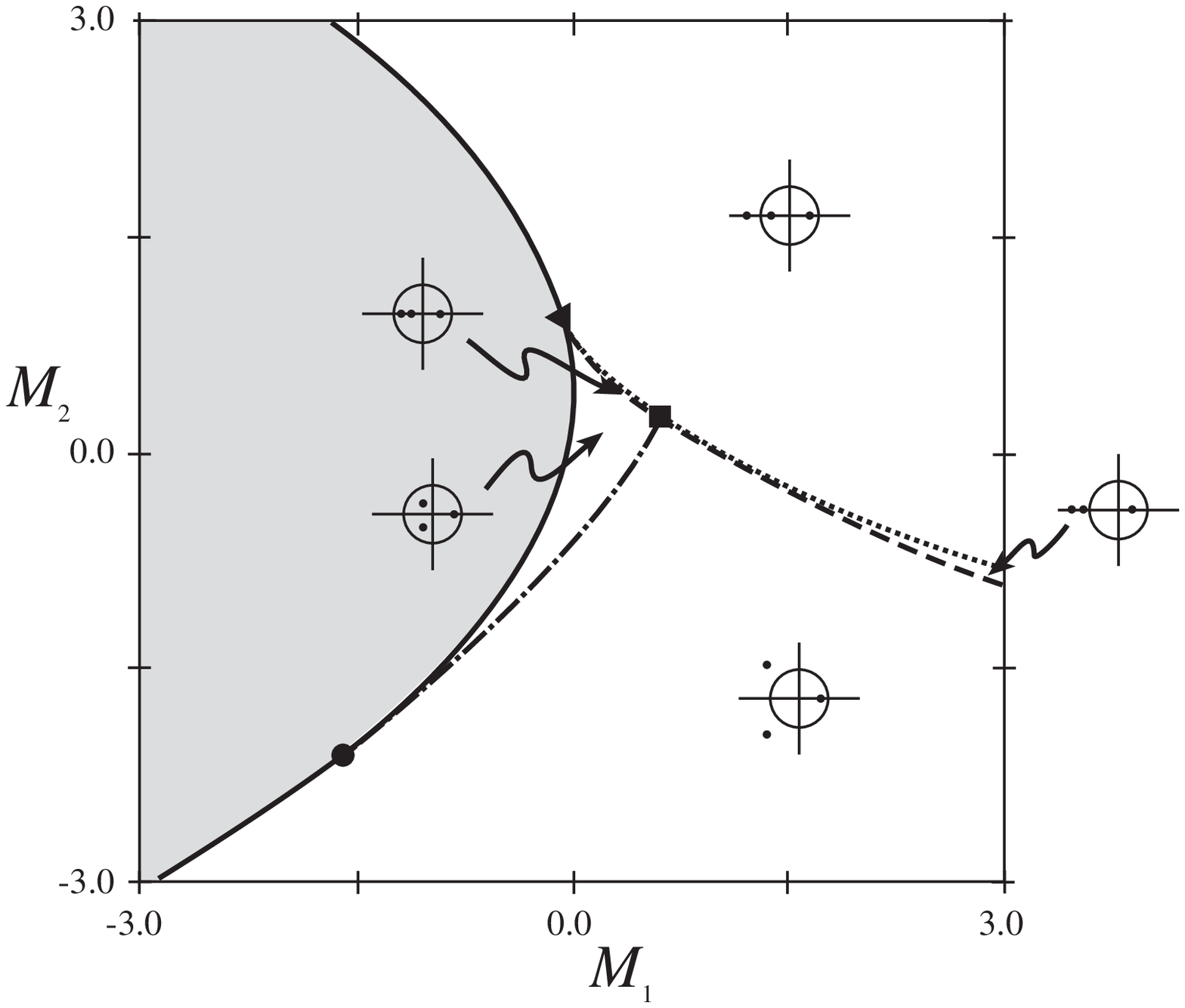}{Bifurcation curves of the map
\Eq{eq:3HM1} for the fixed point $P_+$ when $B=0.6$.  Fixed points
do not exist in the shaded region.  The solid curve is the
saddle-node $L^{sn}$, the dotted curve is the period doubling,
$L^{pd}$, and dot-dash curve is the Hopf bifurcation,
$L^{\varphi}$.  The fourth curve (dashed) corresponds to the
existence of a double eigenvalue.  There are five corresponding
regions, and a representative configuration of the multipliers in
the complex-$\lambda$-plane is shown for each.  The
codimension-two bifurcations are shown as the ``square"
(period-doubling-Hopf), ``triangle" (saddle-node-period doubling)
and ```circle" (saddle-node-Hopf)
points.}{fig:type21bifB06fp1}{4.5in}

%
%
%
\InsertFig{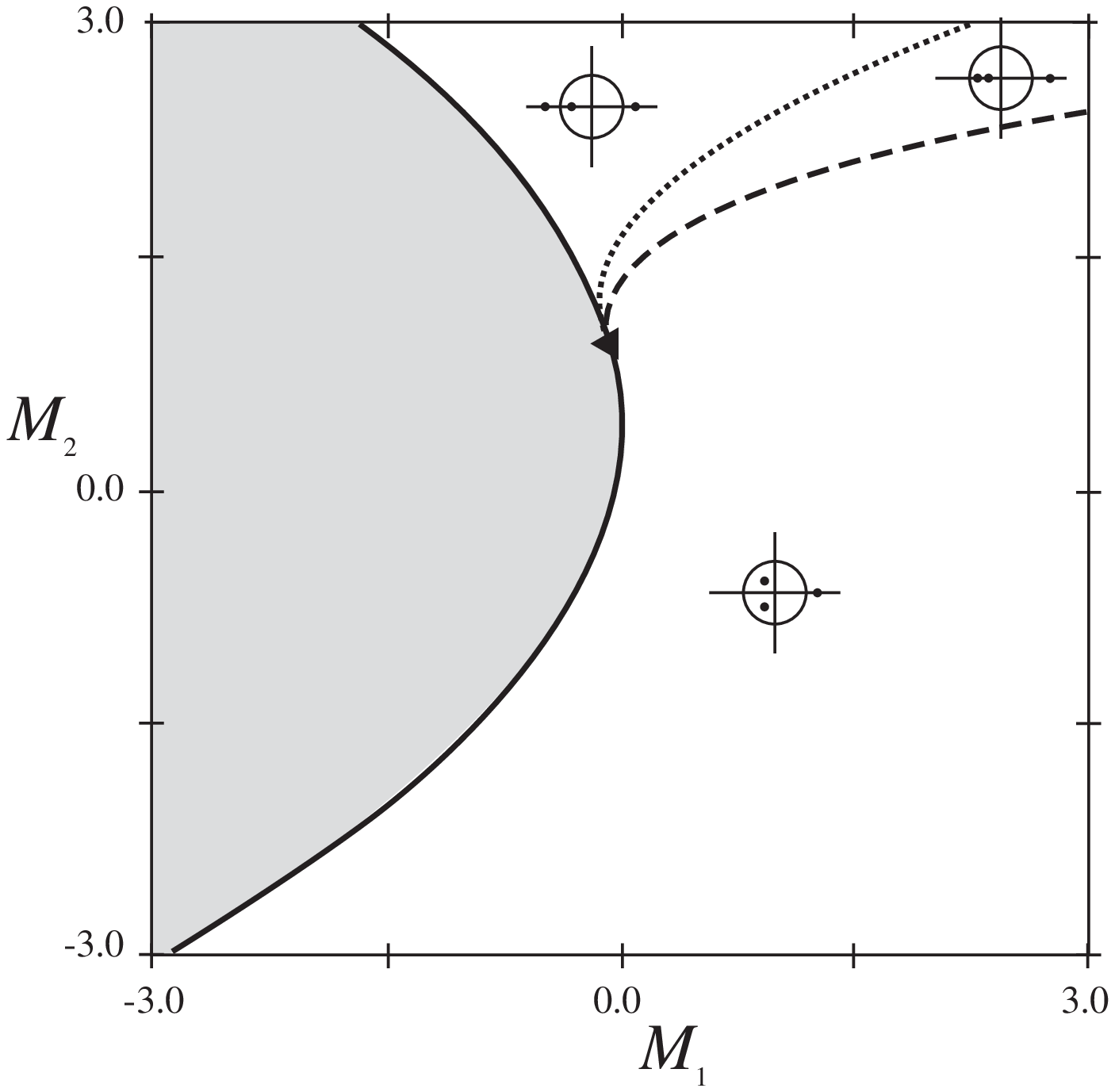}{Bifurcation curves of the map \Eq{eq:3HM2}
in for the fixed point $P_-$ when $B=0.6$.  The curves have the same
meaning as in \Fig{fig:type21bifB06fp1}.  For this fixed point there
is no Hopf bifurcation.}{fig:type21bifB06fp2}{4in}

A similar analysis can be used to derive the bifurcation surfaces for
the map \Eq{eq:3HM2}.  Indeed, the formula for the fixed points, and
consequently the saddle-node surface is the same as for \Eq{eq:3HM1}.
The period-doubling and Hopf surfaces are slightly modified.  An
example of a slice through these surfaces for $B = 0.6$ is shown in
\Fig{fig:type12bifB06fp1} and \Fig{fig:type12bifB06fp2}.
\InsertFig{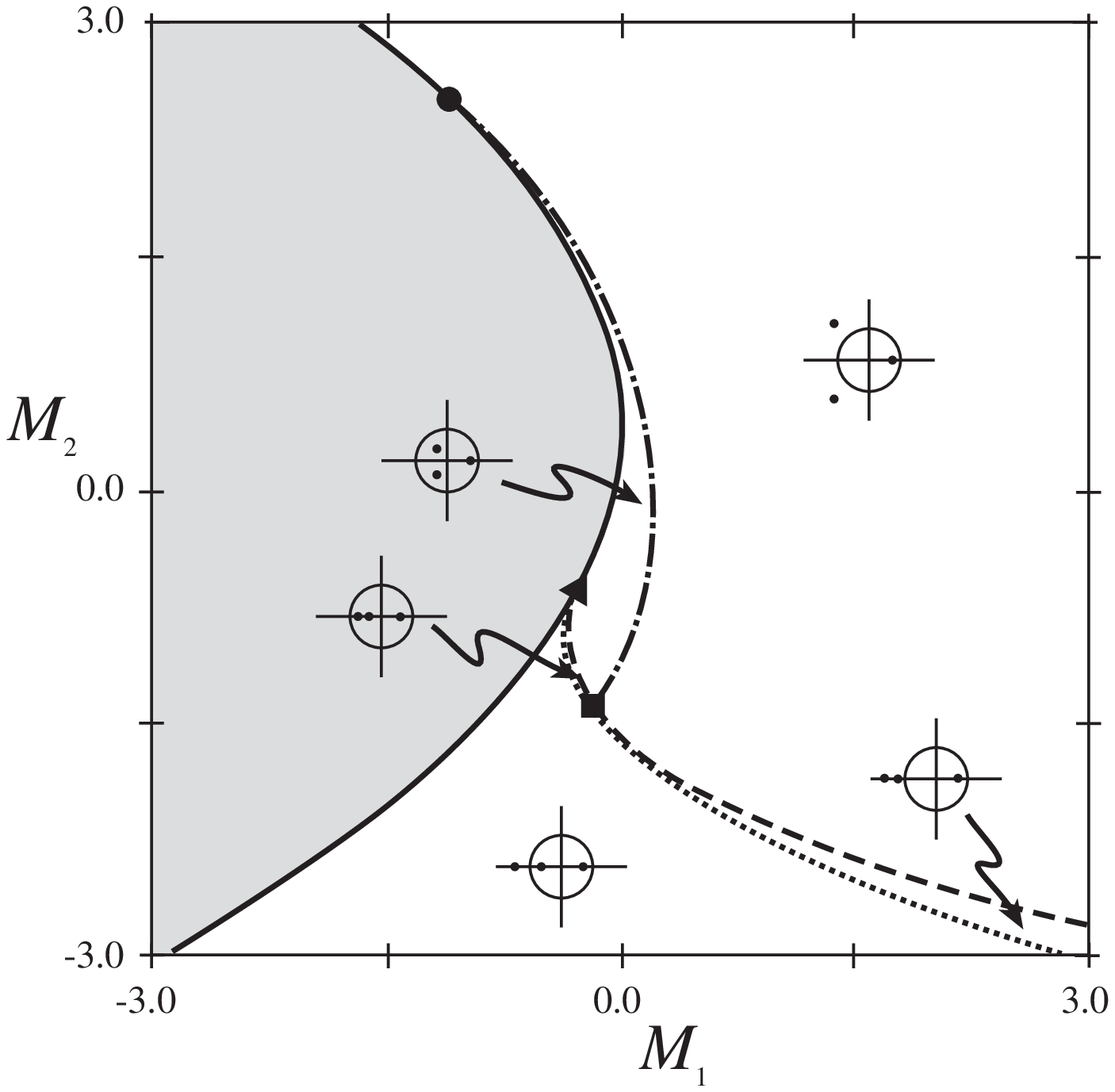}{Bifurcation curves of the map \Eq{eq:3HM2}
for the fixed point $P_+$ when $B=0.6$.  The curves have the same
meaning as in \Fig{fig:type21bifB06fp1}.}{fig:type12bifB06fp1}{4.5in}
\InsertFig{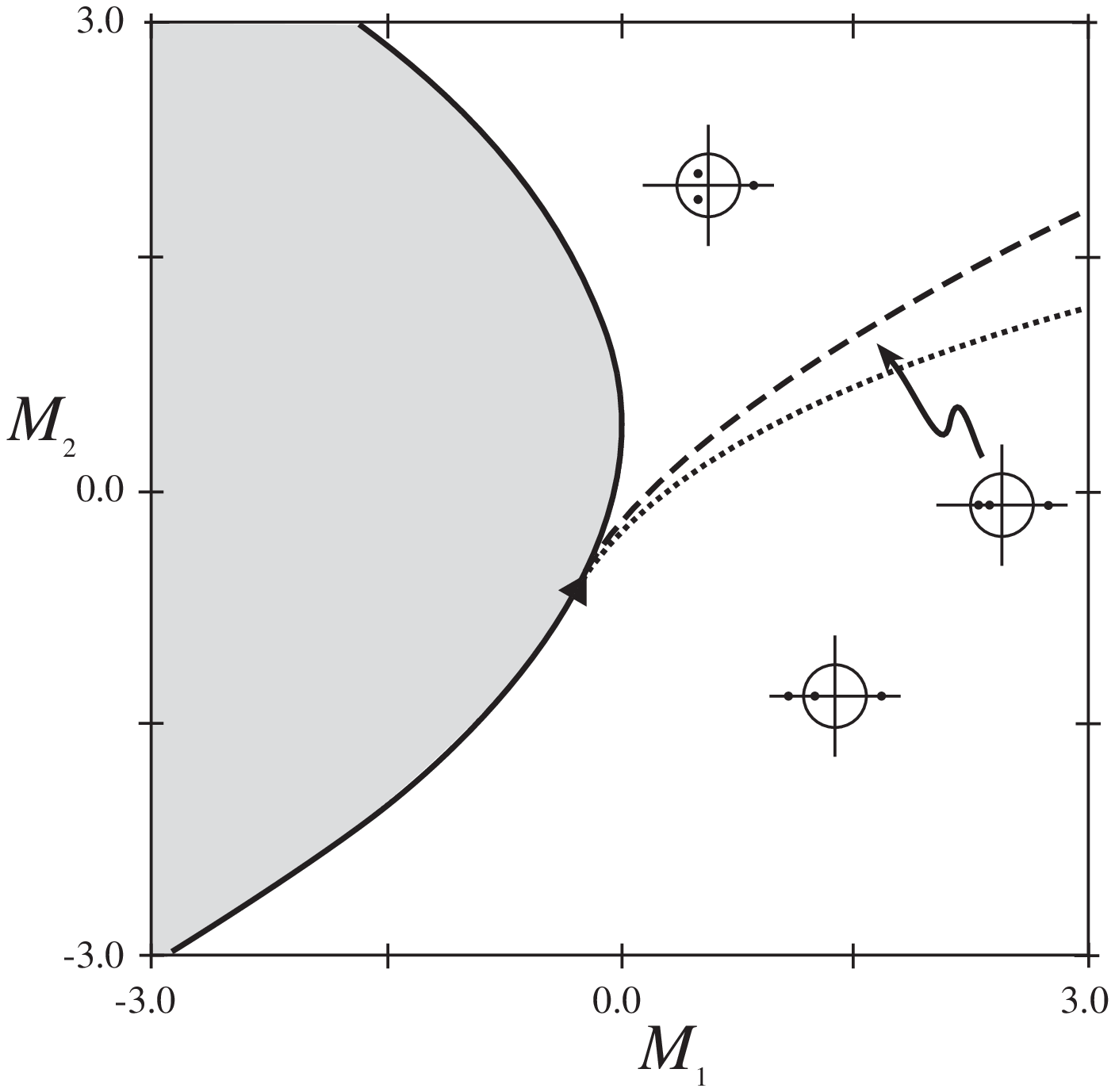}{Bifurcation curves of the map \Eq{eq:3HM1}
in for the fixed point $P_-$ when $B=0.6$.  The curves have the same
meaning as in \Fig{fig:type21bifB06fp1}.  For this fixed point there
is no Hopf bifurcation.}{fig:type12bifB06fp2}{4in}

\section{Wild-hyperbolic Lorenz attractors: Proof of \Th{tm:2}}\label{sec:attractors}
We will now focus on the point $P_+$ and expand around the
codimension-three bifurcation point where the multipliers are
$(-1,-1,1)$, that is about the point $(M_1, M_2, B) = (-\frac14, 1,
1)$ where $z^* = 1/2$.  Note that on the parameter slice $B=1$, the
Hopf curve \Eq{eq:hopf} coincides with the saddle-node curve
\Eq{eq:sn} so that there is no Hopf bifurcation.  Moreover the
codimension-two points in \Fig{fig:type21bifB06fp1} corresponding to
the multipliers $(-1,-1,B)$ (square) and $(-1,1,B)$ (triangle) collide
at the codimension-three point.  Otherwise the bifurcation diagram for
$P_+$ looks similar the slice in the figure.

Our goal is to prove \Th{tm:2}. This can be done with the aid of the following lemma.

\begin{lm}\label{smlem}
  In a region of the parameters $(M_1, M_2, B)$ near the point
  $(-1/4, 1,1)$ the second iterate of \Eq{eq:3HM1} in
  some neighborhood of $P_+$ is approximately the time-one map of the
  the three-dimensional flow
  \begin{equation}  \label{=sm}
      \begin{array}{l}
      \dot x=y \;,\\
      \dot y=x(1-z)-\lambda y \;,\\
      \dot z=-\alpha z+x^2 \;,
    \end{array}
  \end{equation}
  up to arbitrary small correction terms. Moreover the
  coordinates $(x,y,z)$ and parameters $\lambda$ and $\alpha$ can
  take arbitrary finite values.
\end{lm}

\begin{proof}
Shifting coordinates $x\to x-z_+,y\to y-z_+,z\to z-z_+$ brings
map \Eq{eq:3HM1}
to the form
\begin{equation}\label{Hen1}
\begin{split}
    \bar x=y,\;\; \bar y=z,\;\;
    \bar z=Bx+M_2y - 2z_+ z-z^2
\end{split}
\end{equation}
Introduce new (small) parameters $\eps_1,\eps_2,\eps_3$
by the following formulas
$$
\eps_1 = 1 - B,\; \eps_2 = 1 - M_2,\;
\eps_3 = 2z_+ -1.
$$
Note that
$$
  \eps_1 >0,\; \eps_1 + \eps_2 + \eps_3 = \sqrt{\Delta} >0.
$$
The first inequality means that $B<1$, and the second is fulfilled
automatically when the fixed points exists, e.g. to the right of the
$L^{sn}$ curve in \Fig{fig:type21bifB06fp1}.  We have also the
following relation
$$
    1 + 4M_1 = \eps_3^2 +2(1+\eps_3)(\eps_1 + \eps_2)
$$
between the new parameters and the parameter $M_1$.
Next, we bring map \Eq{Hen1} to the following form
\begin{equation}\label{res1}
  \begin{array}{l}
    \bar u_1=-u_1+u_2,\\
    \displaystyle \bar u_2=\alpha_1u_1+(-1+\alpha_2)u_2+
    u_1u_3-\frac 32u_2u_3 + \cO(\|u\|^3),\\
    \displaystyle \bar u_3=
    (1+\alpha_3)u_3-\frac 14u_1^2-\frac 14u_3^2 + \cO(\|u\|^3),
  \end{array}
\end{equation}
where
$$
  2\alpha_1=\eps_1-\eps_2+\eps_3,\:\,
  4\alpha_2=\eps_1+\eps_2-3\eps_3,\:\,
  4\alpha_3=- (\eps_1+\eps_2+\eps_3).
$$
Note that the Jacobian of the map \Eq{res1} is close to the Jordan
normal form for the multipliers $(-1,-1,1)$ when
$\alpha_1,\alpha_2,\alpha_3$ are small.  The first equation of
\Eq{res1} does not contain any high order terms, and in the second and
third components, we take into account only quadratic terms.
Moreover, these quadratic terms have been further simplified by
standard normal form analysis, so that only the resonant ones remain.

The second power of the map \Eq{res1} can be embedded into a flow.
Namely, using the Picard iteration procedure (of the second order), we
can see that the map \Eq{res1} coincides, up to terms of order
$\cO(\|u\|^3)$, with the time-one map of the following flow
\begin{equation}\label{fl1}
  \begin{array}{l}
    \displaystyle \dot u_1=-2u_2-u_1u_3+\frac 56u_2u_3\\
    \displaystyle \dot u_2=\beta_1u_1-\beta_2u_2-2u_1u_3+2u_2u_3\\
    \displaystyle \dot u_3=\beta_3u_3-\frac 12u_1^2-\frac 1{12}u_2^2-\frac 12u_3^2
    -\frac 12u_1u_2.\\
  \end{array}
\end{equation}

After a smooth change of coordinates of the form
$$
  v_1 = u_1,\; v_2 = \dot u_1, \; v_3 = u_3 + \cO(\|u\|^2) \;,
$$
equations \Eq{fl1} are written as follows
\begin{equation}\label{flow}
  \begin{array}{l}
    \dot v_1=v_2\\
    \dot v_2=\bar\mu v_1-\bar\lambda v_2-av_1v_3-a_1v_2v_3\\
    \dot v_3=-\bar\alpha+v_3^2+b(v_1^2+v_2^2),
    \end{array}
\end{equation}
where
$$
  \displaystyle a=8,\:\,a_1=2,\:\, b=\frac 14,\:\,\bar\mu=-\eps_1-3\eps_2-\eps_3,\\\
  \displaystyle \bar\lambda=\frac{\eps_1+3\eps_2-3\eps_3}2,\:\,
  \bar\alpha=\frac{(\eps_1+\eps_2+\eps_3)^2}{16}.
$$
Flow \Eq{flow} is the same that was considered \cite{SST93} and, in
our case we have also that $ab>0$. Then, we can introduce new time $\tau$
and
new rescaled coordinates $(x,y,z)$, by formulas
\begin{equation}\label{newt}
  \tau = s t, \;\;{\rm where}\;\;
  s^2=\bar\mu+a\sqrt{\bar\alpha}=\eps_1-\eps_2+\eps_3,
\end{equation}
\begin{equation}\label{newx}
  \displaystyle v_1\to x\sqrt{\frac{s^3}{ab}},\:\,
  v_2\to y\tau\sqrt{\frac{s^3}{ab}},\:\,
  v_3\to -\sqrt{\bar\alpha}+\frac{s^2}az,\:\,
\end{equation}
such that system \Eq{flow} is reduced to the following one
\begin{equation}\label{sm}
  \begin{array}{l}
    \dot x=y,\\
    \dot y=x(1-z)-\lambda y + \cO(s),\\
    \dot z=-\alpha z+x^2 + \cO(s),
  \end{array}
\end{equation}
where
\begin{equation}\label{scale}
  \begin{array}{l}
    \displaystyle \alpha=\frac{2\sqrt{\bar\alpha}}\tau=
    \frac{\eps_1+\eps_2+\eps_3}{2\sqrt{\eps_1-\eps_2+\eps_3}},\;\;
    \lambda=\frac {\bar\lambda}\tau=
    \frac{\eps_1+3\eps_2-3\eps_3}{2\sqrt{\eps_1-\eps_2+\eps_3}}.
  \end{array}
\end{equation}
Since values of $s$ can be arbitrarily small, it implies that the rescaled
coordinates $(x,y,z)$ and new parameters $\lambda$ and $\alpha$ can
take arbitrarily finite values.
\end{proof}

Note that if we omit the $\cO(s)$ terms in the system \Eq{sm}, we
obtain system \Eq{=sm}, which is the well-known, Shimizu-Morioka
model.  This system was intensively studied in
\cite{ShA86},\cite{ShA93} where, in particular, the existence of
the Lorenz attractor was proved for some domain of the parameters.
For example, the domain of the existence of the Lorenz attractor
in Shimizu-Morioka model is known to be contained in a rectangle
$0.7<\lambda<1.5, \:\,0.1<\alpha<0.6$ (\cite{SST93}, fig.  13).
Using this fact we can pointed out the following boundaries of
values of the parameters $(\eps_1,\eps_2,\eps_3)$:
\begin{equation}
    \begin{array}{l}
    \eps_1>0 \;,\\
    \eps_1+\eps_2+\eps_3>0\;,\\
    \eps_1-\eps_2+\eps_3>0\;,\\
    \displaystyle 0.7<\frac{\eps_1+3\eps_2-3\eps_3}{2\sqrt{\eps_1-\eps_2+\eps_3}}<1.5\;,\\
    \displaystyle 0.1<\frac{\eps_1+\eps_2+\eps_3}
    {2\sqrt{\eps_1-\eps_2+\eps_3}}<0.6\;,
    \end{array}
\end{equation}
(and, respectively, of the parameters $(M_1,M_2,B)$) such that in
the corresponding domain of the initial parameters $(M_1,M_2,B)$
our 3D H\'enon map has a wild-hyperbolic Lorenz-type attractor.
This completes the proof of the main theorem.\\~\\

\section{Quadratic Volume-Preserving Mappings: Proof of \Th{tm:4}}\label{sec:quadratic}

Suppose that $f: \R^3 \rightarrow \R^3$ is a quadratic Cremona map,
i.e., is a diffeomorphism with constant Jacobian $\det(Df) = B$.  Then
it is easy to see that without loss of generality we can write it in
the form
\[
    f = S \circ L \circ (id + Q) \;.
\]
where $S(x,y,z) = (B^{1/3}x,B^{1/3}y,B^{1/3}z)$ is the diagonal
scaling map with $\det(S) = B$, $L$ is an affine volume-preserving
map, $Q$ is a vector of quadratic forms so that the map $x+Q(x)$ is a
volume-preserving \textit{quadratic shear} \cite{LM98}.

As in \Th{tm:4} we now assume that the map $f$ has a polynomial
inverse (i.e., it is an element of the affine Cremona group) and that
the inverse is also a quadratic map.  The same property then will also
hold for the volume-preserving map $g = S^{-1}\circ f$.  It was shown
in \cite{LM98} that maps of the form $g$ can be transformed, by an
affine change of variables, to one of the three following forms:
\begin{equation*}
    \begin{split}
        g_1(x,y,z) &= (y,z, M_1+M_2z+M_3 y + x + q(y,z)) \;, \\
        g_2(x,y,z) &= (x_0 + \frac{1}{M_1}x, y_0 - M_1 z, z_0+ M_2z + y + q(x,z)) \;, \\
        g_3(x,y,z) &= (x_0 + y + M_1 x, y_0 -\frac{1}{M_2}x, z_0 + M_2 z + q(x,y)) \;,
    \end{split}
\end{equation*}
where $q(x,y) = ax^2 + b xy + cy^2$ is a scalar quadratic form.

Consequently $g = A \circ g_i \circ A^{-1}$ for some affine
transformation $A$, and $i = 1$, $2$, or $3$.  Using the same
transformation on $f$ gives
\[
    \tilde f = A^{-1} \circ f \circ A = (A^{-1} \circ S \circ A) \circ g_i \;.
\]
But the scaling map $S$ commutes with $A$, so
\[
    \tilde f = S \circ g_i \;.
\]
Finally we can apply an additional scaling transformation to the
variables and parameters to transform $\tilde f$ to one of the
three forms
\begin{equation}\label{eq:qforms}
    \begin{split}
        \tilde f_1(x,y,z) &= (y,z, M_1+M_2z+M_3 y + Bx + q(y,z)) \;, \\
        \tilde f_2(x,y,z) &= (x_0 + \frac{1}{M_1}x, y_0 - M_1 z, z_0+ M_2z + By + q(x,z)) \;, \\
        \tilde f_3(x,y,z) &= (x_0 + y + M_1 x, y_0 -\frac{1}{M_2}x, z_0 + BM_2 z + q(x,y)) \;,
    \end{split}
\end{equation}
where the parameters have been rescaled.

The dynamics of $\tilde f_2$ and $\tilde f_3$ are trivial. Indeed
for $\tilde f_2$ the $x$ dynamics decouples from the remaining
dynamics, and for $\tilde f_3$ the $(x,y)$ dynamics is linear and
decouples from the $z$ dynamics.

Thus the only dynamically interesting case is $\tilde f_1$.  Since
this map is the same as \Eq{eq:3HM3}, we have proved \Th{tm:4}.

An example of this analysis consider the ACT-map
(Arneodo-Coullet-Tresser map) which was introduced by Arneodo,
Coullet and Tresser in their study of strange attractors in a
family of differential equations on $\R^3$ with homoclinic points
of Shilnikov type \cite{BSDU85}.
\begin{equation}\label{eq:ACT}
    \begin{array}{l}
        \bar x = ax - b(y-z),\\
        \bar y = bx + a(y-z), \\
        \bar z = cx - d x^k + ez,
    \end{array}
\end{equation}
where $a,b,c,d,e$ are real parameters with $J = e(a^2+b^2) \neq 0$,
$bd\neq 0$, and $k>1$ an integer.  The ACT-map always has a fixed
point at the origin.  Nevertheless, when $k=2$ it is a quadratic
family with a quadratic inverse, so it follows from \Th{tm:4} that it
is affinely conjugate to \Eq{eq:3HM3} with $M_1=0$.

To see this explicitly define new coordinates using the transformation
$$
    X = \frac{ax+by}{a^2+b^2}\;,\; Y = x\;,\; Z = ax - b(y-z) \;,
$$
then, the map is rewritten as
$$
    \begin{array}{l}
        \bar X = Y,\\
        \bar Y = Z, \\
        \bar Z = BX +S_1Y + S_2 Z - bd Y^k
    \end{array}
$$
where where $B = e(a^2+b^2), S_1 = - (a^2+b^2+ cb+2ae), S_2 = 2a+e$.
For the case $k=2$ we can rescale the coordinates by a factor $(bd)^{-1}$ to set
the coefficient of $Y^2$ to one. Thus the map is written in the promised form \Eq{eq:3HM3}.

Note that if we do one additional transformation and shift the variables by $S_1/2$ we can also obtain the standard form \Eq{eq:3HM2}. Here we define  $M_1= \frac14 S_1(2B+2S_2+S_1-2)$, and $M_2 = S_2$. The inequality
$$
    \Delta = (1-B-M_2)^2 + 4M_1 = (B+S_1+S_2-1)^2 \geq 0
$$
holds guaranteeing the existence of a fixed point for the map.

\section{Conclusions}
We have seen that many phenomena in the dynamics of
three-dimensional H\'enon maps are markedly different from the
possible phenomena predicted by our knowledge of two-dimensional
dynamics.  Primarily, we have seen these new phenomena when the
Jacobian, $B$, is close to $\pm 1$.  Two natural problems appear
here, regarding the understanding of bifurcation scenarios that
lead to chaotic three-dimensional dynamics. The first is to
understand ``how 2D dynamics is transformed into 3D dynamics" upon
the variation of $B$ from values near zero to those close to $\pm
1$.  The second problem---which is, in a sense, the inverse of the
first one---is to understand ``how 3D volume-preserving dynamics
is transformed into dissipative (and chaotic) dynamics" as $B$
moves away from $\pm 1$.  In the current paper (and also in
\cite{GOST05}), some partial results related to the second problem
have been obtained; however, the first problem is---as of
yet---unstudied.

However, our analysis implies that these global problems have no
unified, qualitative solution.  One general conclusion is that the
stable dynamics of the H\'enon maps of the first and second kind,
\Eq{eq:3HM1} and \Eq{eq:3HM2}, have unmistakably different characters.
The bifurcations in \Eq{eq:3HM1} and \Eq{eq:3HM2} both lead eventually
to the formation of three-dimensional versions of Smale horseshoes.
However, these have different characters: the horseshoes of
\Eq{eq:3HM1} contain type $(2,1)$ saddles of, while those of
\Eq{eq:3HM2} contain type $(1,2)$ saddles.  Thus, a third global
problem appears: to understand ``how the character of the attractors
changes as one changes type of nonlinearity in 3D H\'enon families."
In this case, it seems that using the family \Eq{eq:3HM3} with the
parameters $(a,b,c)$ determining the type of nonlinearity may be very
profitable.

While the problems posed above are of general interest, there are also
many partial problems relating directly to the dynamics of quadratic
maps.  Some of these may also be solved in the the near future:
\begin{enumerate}
    \item Does the H\'enon map of the second kind, \Eq{eq:3HM2}, have
        a large scale, wild-hyperbolic attractor?
    \item What are the dynamical properties of the local normal form
        for \Eq{eq:3HM2} near fixed points with three unit multipliers?

    This is related to the "reverse-Shimitsu-Marioka" system since
    \Eq{eq:3HM2} is the inverse of \Eq{eq:3HM1}. Thus we will obtain
    the model \Eq{=sm}, with $t$ replaced by $-t$.
    Consequently, the question is to determine the properties of the stable dynamics of
    this system.

    \item What are main bifurcations scenarios the 3D H\'enon map
    \Eq{eq:3HM1} that
        lead from simple dynamics to a wild-hyperbolic, Lorenz attractor?

    There are two distinct, but useful approaches: (a) to fix $B$ and consider the
    corresponding two-parameter family; (b) to consider $B$ as the
    main governing parameter and to move for given values of $(M_1,M_2)$
    from small values of $B$ to values close to $\pm 1$.

     \item What role is played by the codimension-two bifurcations
        (recall \Sec{sec:bifurcations}) in the formation of attracting
        invariant sets?

    The analysis of codimension-one bifurcations alone gives us only
    a representation of the main hyperbolic elements of the dynamics.
    A detail analysis of codimension-three
    bifurcations (as in \Sec{sec:bifurcations}) allows us to find in
    local normal forms (using the flow embedding approach) and verify the existence of
    small, wild-hyperbolic attractors. Analogous methods can be applied to
    the study of codimension-two bifurcations. We believe that this approach will
    provide an understanding of principal mechanisms for the creation of wild-hyperbolic
    attractors.
\end{enumerate}

In summary, three-dimensional H\'enon maps have an abundance of new
dynamical behavior which, it seems, has an important place in the
theory of multidimensional dynamics as whole.  We believe that they
form a natural direction for a program of investigation similar to the
well developed programs in one and two-dimensional dynamics.  This is
especially true since these maps cover (up to the affine conjugacy)
all possible quadratic three-dimensional automorphisms.


\end{document}